\documentclass{article}
\usepackage{spconf,amsmath,graphicx,hyperref}
\usepackage{booktabs}
\usepackage{comment}
 \usepackage{multirow}

\usepackage[activate]{microtype}
\title{Perceptible or Not? Diagnosing Passive Fingerprints for Speech Deepfake Attribution}
\name{Yupei Li$^{1,4}$, Qiyang Sun$^{1}$, Emmanouil Benetos$^{2}$, Berrak Sisman$^{3}$, Bj\"orn Schuller$^{1,4}$}

\address{
$^{1}$Imperial College London,
$^{2}$Queen Mary University of London, \\
$^{3}$Johns Hopkins University,
$^{4}$Technical University of Munich
}
\begin{document}
\ninept
\maketitle
\begin{abstract}
Passive fingerprints (intrinsic traces naturally left by generators) have been shown to enable attribution in speech deepfake detection, yet their persistence, reproducibility, and content-independence remain unverified. Moreover, no prior work distinguishes perceptible from imperceptible fingerprints, although the two have very different implications for attribution reliability. Perceptible fingerprints, such as emotional expression, are shaped by perceptual quality objectives and may change across model updates, whereas imperceptible fingerprints are not explicitly optimised by current training objectives and are rarely considered in existing dataset design or training strategies, as they have limited influence on downstream applications. We therefore propose a Perceptible-Imperceptible Passive-fingerprint Diagnostic Protocol (PIPDP) to define and separately analyze these two fingerprint types. PIPDP comprises three complementary analyses: multi-evidence fingerprint verification through residual-energy, reproducibility, and saliency analyses, perceptually transparent perturbations preserving audio quality, and prompt-driven emotion change that modifies perceptible fingerprints without model retraining. Experiments across ten speech generators and three attribution detectors show that imperceptible fingerprints provide persistent attribution cues. Perceptually transparent perturbations reduce attribution accuracy by up to 48.2\% on HiggsAudioV3, whereas emotion-driven changes leave attribution largely unchanged, with only about a 1.0\% accuracy variation across emotions on CosyVoice2 using w2v-bert-MLP. These results suggest that imperceptible fingerprints are more reliable for trustworthy attribution.
\end{abstract}
\section{Introduction}
\label{sec:intro}
As neural speech generators advance, source attribution is increasingly important for forensic analysis, identifying the generator of synthetic speech~\cite{Klein_2024}. Attributors typically learn discriminative features~\cite{chen2025towards} from audio signals containing generator-specific fingerprints~\cite{chuyuan-etal-2024-distinguishing}.Such fingerprints come in two forms: active watermarks (e.g., SynthID\footnote{\url{https://deepmind.google/models/synthid/}}, AudioSeal~\cite{pmlr-v235-san-roman24a}), deliberately embedded for copyright protection, and passive fingerprints, unintended traces left by generators. Unlike active watermarks designed for explicit traceability, passive fingerprints reveal intrinsic generator characteristics and support broader real-world applications, such as forensic analysis \cite{cassia2025deepfake}.


Previous definition of fingerprint is defined as the set of deviations of generated samples from the true data \cite{song2024manifpt}, which is not intuitive for a human observer. Unlike discrete modalities such as text, audio is a continuous signal with a substantial portion of information beyond direct human perception. Additionally, as future generator releases optimize for perceptual quality, any fingerprint tied to that quality is likely to be changed, while one decoupled from it may persist across versions \cite{camara2025neural}. We therefore distinguish passive fingerprints into two categories: \textbf{perceptible fingerprints}, which are human-audible, and \textbf{imperceptible fingerprints}, which are not consciously perceived but still preserve source-related information. Since imperceptible fingerprints do not affect perceived audio quality, they are rarely optimized explicitly, resulting in limited supervised datasets and systematic analyses, while only a few preliminary studies have explored their potential for attribution~\cite{lin2025towards}.

Despite these observations, passive fingerprints remain poorly understood. Prior work has examined fingerprint robustness under noise, compression, and adversarial perturbation \cite{pizarro2024lightweight,jin2026rvcbenchbenchmarkingrobustnessvoice}, and questioned whether the fingerprints are robust~\cite{fi18070344}. However, these studies focus on robustness to external degradations rather than the intrinsic properties of fingerprints, namely persistence, reproducibility, and content independence. These properties remain largely unexplored. To address these gaps, we propose a Perceptible-Imperceptible Passive-Fingerprint Diagnostic Protocol (PIPDP) to investigate the reliability of imperceptible fingerprints. PIPDP systematically diagnoses fingerprints through consistency analysis, repeatability testing, perceptually controlled perturbations, and generation-condition ablations. 



In summary, we make three \textbf{contributions}. First, we propose a conceptual framework distinguishing perceptible from imperceptible passive fingerprints. Second, we design a lightweight, model-agnostic diagnostic protocol, and training-free input ablations that reveal imperceptible fingerprints. Third, we performed experiments on ten speech generators and three cross-architecture detectors, showing that imperceptible fingerprints provide more persistent and reliable attribution signals than perceptible fingerprints.

\section{Method: Perceptible-Imperceptible Passive-fingerprint Diagnostic Protocol}
\begin{figure}[h]
    \centering
    \includegraphics[width=\linewidth]{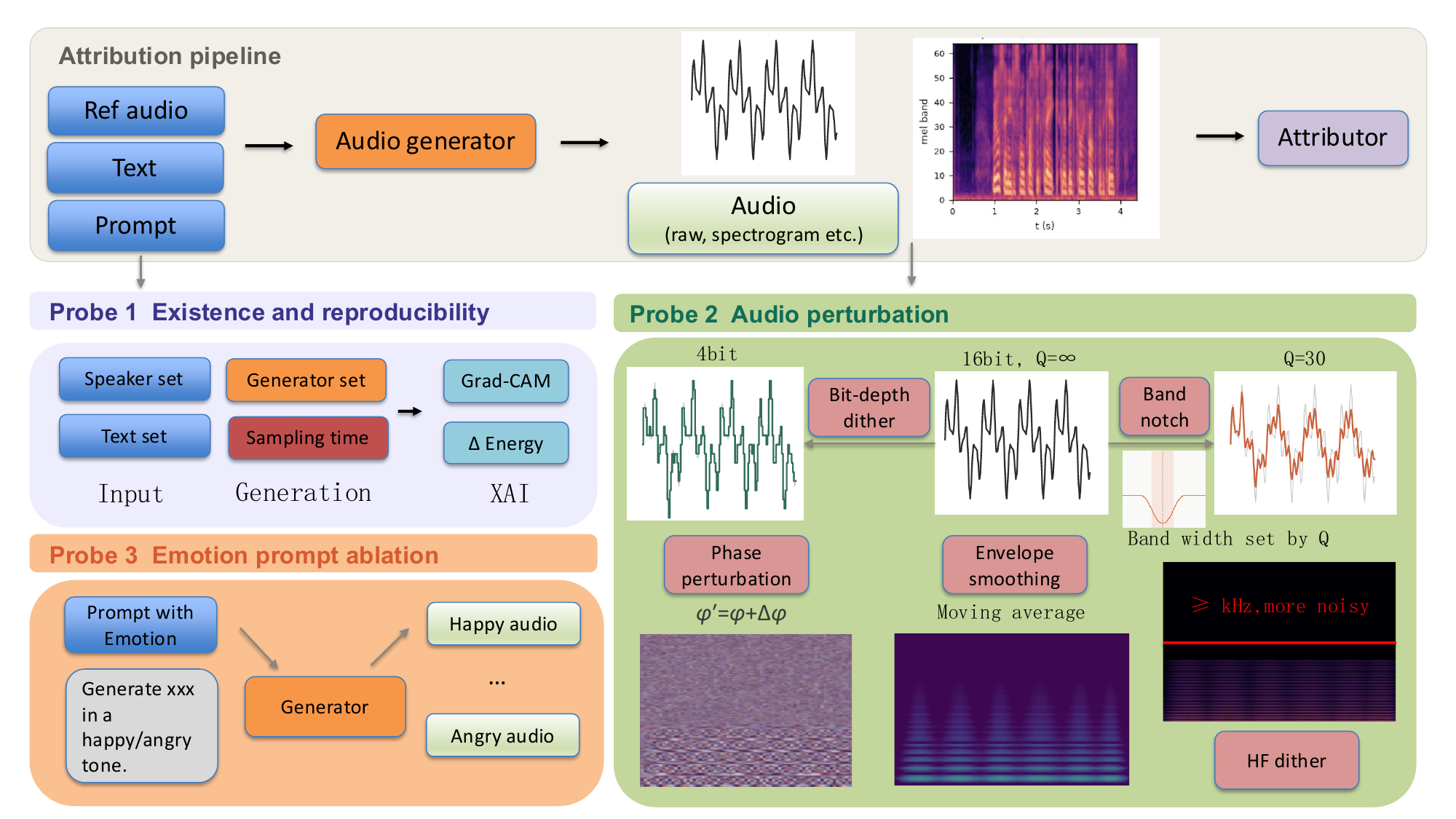}
    \caption{PIPDP methodology. Given generated speech from different generators, PIPDP evaluates attribution using cross-architecture attributors in a closed-set setting. It diagnoses fingerprints through three probes: (1) existence and reproducibility, analyzing residual energy and attribution explanations under speaker, text, and sampling variations; (2) audio perturbation, testing whether attribution relies on imperceptible fingerprints; and (3) emotion ablation, modifying perceptible attributes through prompts to evaluate their stability.
}
    \label{fig:pipdp}
    \vspace{-0.7cm}
\end{figure}

PIPDP characterizes passive fingerprints along the axis of human perceptibility. As shown in Figure~\ref{fig:pipdp}, it operates on a common attribution pipeline, where speech is generated from controlled inputs and evaluated by attributors. The three probes are designed to cover the key diagnostic dimensions of passive fingerprints, which together are sufficient to address the our central claims to \textbf{analysis the reliability of imperceptible fingerprints}. Probe 1 establishes that attribution signals correspond to persistent and reproducible generator-specific traces. Probe 2 examines whether these reliable fingerprints are imperceptible. Probe 3 then evaluates the stability and reliability of perceptible fingerprints. We quantify perturbation effects using accuracy drop relative to correctly attributed samples and probability drop of the true class. Decrease in these metrics indicates that the perturbed fingerprint components contribute to attribution.

\subsection{Attribution pipeline}
For a comprehensive diagnostic analysis of deepfake attribution, we select ten speech generators spanning three functional categories: vocoders (HiFiGAN \cite{kong2020hifi}, MelGAN \cite{kumar2019melgan}), text-to-speech (TTS), and voice conversion (VC). The TTS and VC systems further span three architectural paradigms: classical synthesis (GPT-SoVITS\footnote{\url{https://github.com/RVC-Boss/GPT-SoVITS}}, FreeVC \cite{li2023freevc}, kNN-VC \cite{baas2023knnvc}), LLM (large language model)-based models (CosyVoice2 \cite{du2024cosyvoice}, XTTS-v2 \cite{casanova2024xtts}, Qwen3-TTS \cite{hu2026qwen3}, HiggsAudioV3\footnote{\url{https://huggingface.co/bosonai/higgs-audio-v3-tts-4b}}), and diffusion models (F5-TTS \cite{chen2025f5}). 

We randomly sample 2,500 utterances from ten speakers (1,800/450/250 for training, in-domain testing, and OOD testing), and synthesize each with all ten generators, yielding 25,000 utterances. We additionally construct another OOD test set by randomly selecting 10 speakers from the ESD dataset \cite{zhou2021emotional}, each contributing 100 English scripts. The data and code are provided\footnote{https://anonymous.4open.science/r/Perceptible-or-Not-3FCB/}. We adopt a closed-set 10-way attribution setting to diagnose fingerprint existence and attribution dependence as a representative, as our conclusions are independent of whether attribution is performed in closed- or open-set settings. To avoid classifier-specific bias, we apply three cross-architecture attributors: WavLM \cite{chen2022wavlm}-AASIST \cite{jung2022aasist}, RawNet2 \cite{tak2021end}, and w2v-bert \cite{chung2021w2v}-MLP. All generators produce 16 kHz speech, with reference audio and target texts from the VCTK corpus \cite{yamagishi2019vctk}.

\subsection{Probe 1: Existence and reproducibility}

Prior work establishes fingerprint existence almost exclusively by showing that a classifier or attribution network achieves high accuracy \cite{yan2022audio}, which may reflect classifier ability rather than the generator-specific trace. Residual energy has been proposed as an explicit fingerprint \cite{pizarro2024lightweight} without systematic analysis. \textbf{Probe 1 instead directly shows that generator fingerprints exist, are persistent, and are reproducible}, triangulating across four complementary signals: attribution accuracy, residual-energy patterns, Grad-CAM saliency \cite{selvaraju2020grad}, and human inspection. Specifically, we compute mel-power spectrograms and frequency-wise residual energy against real references following \cite{pizarro2024lightweight}, and evaluate cross-content consistency, cross-generator distinguishability, and reproducibility (for stochastic generators). 

\subsection{Probe 2: Audio perturbation}

\textbf{To determine whether persistent, reproducible fingerprints reside in the perceptible or imperceptible components of the signal, we design five perturbation operators, most of which are intended to be perceptually transparent in Probe 2.} General quality metrics such as DNSMOS \cite{reddy2021dnsmos}, STOI \cite{taal2010short}, and Word Error Rate (WER) \cite{jurafsky2025speech} have been adopted as objective measures of whether a perturbation is perceptible. The aforementioned accuracy and probability drops are used to quantify the reliability of imperceptible fingerprints.

Let $Z(t, f)$ be the short-time Fourier transform (STFT) of $x[n]$, with $t$ indexing time frames and $f$ indexing frequency bins. The magnitude and phase are $|Z(t, f)|$ and $\varphi(t, f)$.

\subsubsection{Phase perturbation (PP)}
We perturb the STFT phase spectrum with small independent uniform noise while preserving the magnitude, as the human auditory system is largely insensitive to phase. The noise scale $\sigma$ is set to 0.2 and 0.4.
$$
Z'(t, f) = |Z(t, f)| \, e^{j \left[ \varphi(t, f)  +\varepsilon(t, f) \right]}, 
\quad \varepsilon(t, f) \sim \mathcal{U}(-\sigma, \sigma).
$$

\subsubsection{High-frequency dithering (HD)}

We inject complex Gaussian noise into the STFT above $f_c =$ 6\,kHz, a region just beyond the primary speech-intelligibility range where perceptual contribution to word recognition is minimal \cite{monson2014perceptual}, with energy proportional to the local high-frequency energy $E_\text{HF}(t)$. The equation is shown below with $r$ set as 20 and 10.
$$Z'(t, f) = Z(t, f)  1[f \geq f_c] \cdot n(t, f), $$
{\fontsize{9}{9}\selectfont
$$\quad n(t, f) \sim \mathcal{CN}\!\left(0, \, r \cdot E_{\text{HF}}(t) \right), E_{\text{HF}}(t) = \frac{1}{|\mathcal{F}_{\text{HF}}|} \sum_{f \geq f_c} |Z(t, f)|^2.
$$

}
\vspace{-0.5cm}
\subsubsection{Envelope smoothing (ES)}
The STFT magnitude is smoothed along the frequency axis using a length-$k$ moving-average filter while preserving the original phase. This suppresses fine-grained spectral variations while retaining coarse spectral structures, targeting fingerprints potentially caused by resolution mismatches between internal representations and STFT analysis.

\subsubsection{Bit-depth dither (BD)}

We requantize the waveform to $b$ (set as 8 and 10) bits precision using triangular probability density function (TPDF) dither. This decorrelates quantization errors from the signal, yielding spectrally flat noise while removing sub-LSB details without tonal artifacts. The operation is defined as:
$$x'[n] = \operatorname{round}\!\left( \frac{x[n] + \xi[n]}{\frac{2}{2^b}} \right) \cdot \frac{2}{2^b}, 
\; \xi[n] \sim \text{TPDF}(-\frac{2}{2^b},\frac{2}{2^b}).
$$

\subsubsection{Band notch (BN)}
We apply a second-order infinite impulse response notch filter following the standard RBJ design \cite{toy2021audioeqcookbook}, with the center frequency $f_0$ randomly selected and the bandwidth controlled by the quality factor $Q$ (40, 20). This selectively attenuates a narrow frequency band to assess whether attribution relies on localized spectral information.

\subsection{Probe 3: Emotion ablation}

The previous probe shows that \emph{imperceptible} fingerprints are influential yet difficult to optimize due to their lack of observability. In contrast, \emph{perceptible} fingerprints are easier to identify and manipulate. \textbf{Probe 3 therefore tests their stability by modifying only perceptible attributes through prompting while keeping generator parameters fixed. A non-trivial attribution drop indicates that perceptible fingerprints are less stable and less reliable.}

For generators supporting emotion prompts (CosyVoice2 and HiggsAudioV3), which have been optimized for emotion control as reported in their technical reports, we generate happy, sad, and angry versions with neutral as the baseline, while fixing the text, speaker reference, model weights, and decoder seed. Only the emotion descriptor is modified minimally in prompts.



\section{Experiments and results}

We train the attributor while freezing the feature extractor (WavLM and w2v-bert) to preserve its original audio representation. The results are reported in Table \ref{tab:backbone}. They indicate that the features including fingerprints are distinguishable by different attribution backbones. Among them, RawNet2 achieves the lowest performance, likely due to its relatively small model capacity. Therefore, we use the remaining two backbones in the subsequent experiments, with w2v-bert-MLP serving as the primary attributor. 

\begin{table}[h]
\centering
\caption{Performance of different attribution backbones under various evaluation settings. Results are reported as Accuracy / Macro-F1. \textit{ID} refers to in domain, \textit{Spk} refers to speaker and \textit{Text} refers to the transcripts content.}
\label{tab:backbone}
\resizebox{\linewidth}{!}{
\begin{tabular}{lcccc}
\toprule
\textbf{Backbone} & \textbf{Test$_{\text{ID}}$} & \textbf{Test$_{\text{SpkOODTextID}}$} & \textbf{Test$_{\text{SpkOOD}}$} & \textbf{Test$_{\text{Full-OOD}}$ (ESD)} \\
\midrule
RawNet2          & 0.7251 / 0.6959 & 0.7220 / 0.7001 & 0.7280 / 0.7007 & 0.5225 / 0.4690 \\
w2v-bert-MLP   & \textbf{0.9973} / \textbf{0.9973} & \textbf{0.9465} / \textbf{0.9443} & \textbf{0.9180} / \textbf{0.9149} & \textbf{0.8571} / \textbf{0.8587} \\
WavLM-AASIST   & 0.9464 / 0.9459 & 0.8720 / 0.8703 & 0.8660 / 0.8627 & 0.6415 / 0.6240 \\
\bottomrule
\end{tabular}

}
\end{table}

\subsection{Existence and reproducibility}
\begin{figure}[h]
    \centering
    \includegraphics[width=\linewidth]{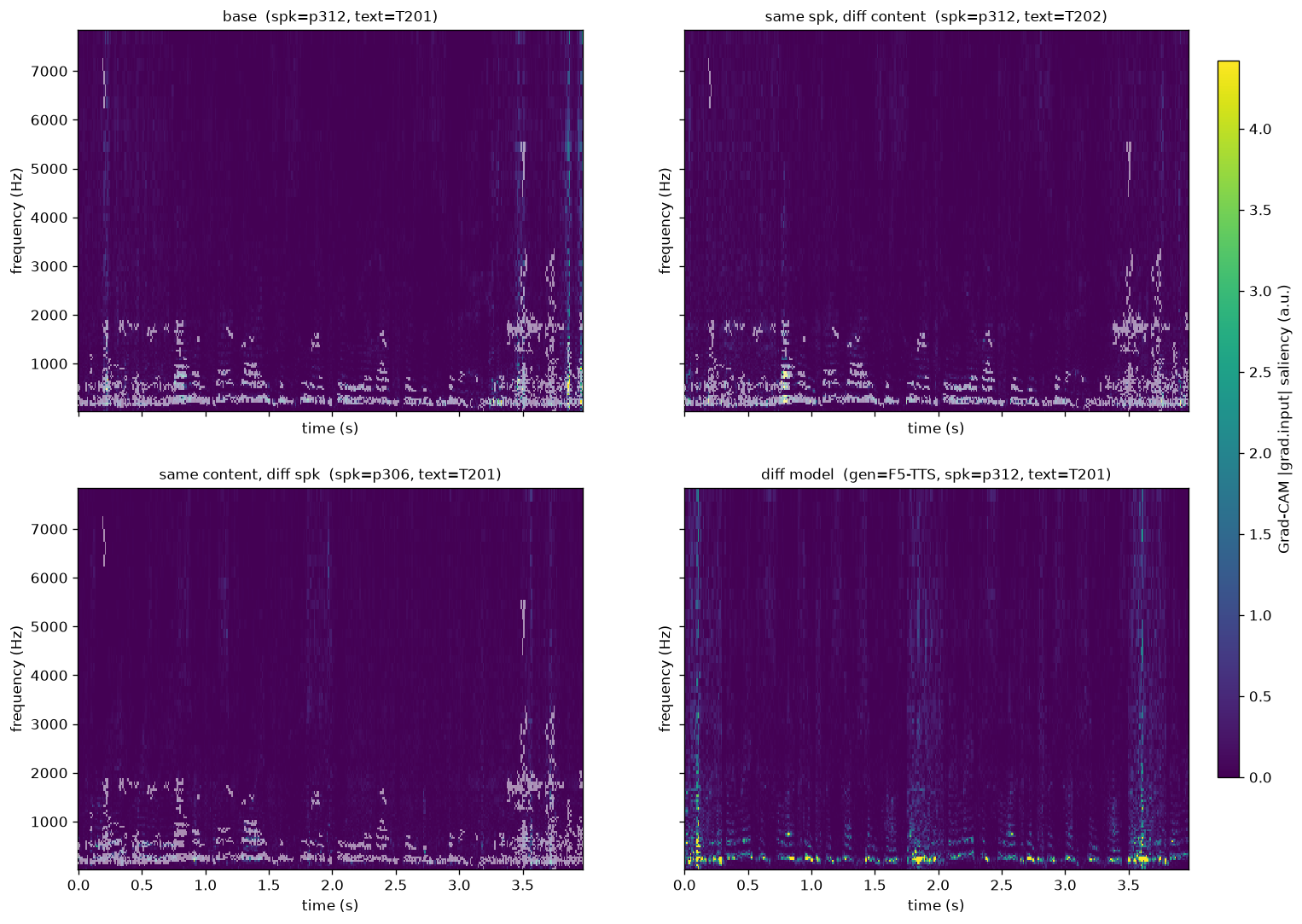}
    \caption{Grad-CAM map for utterances generated by HiFiGAN, with different model using F5-TTS (right bottom). White regions indicate the overlapping salient regions shared across different utterances (top 15\% highest-saliency regions are retained). \textit{spk} means the speaker id.}
    \label{fig:hifigan}
\end{figure}

To visualize whether salient regions persist across content variations, we compute Grad-CAM saliency maps for representative samples from each generator under three conditions: (i) the same speaker with different content, (ii) different speakers with the same content, and (iii) different generators with the same content. Figure~\ref{fig:hifigan} shows one example  in VCTK from HiFiGAN and F5-TTS, correctly classified by the w2v-bert-MLP. The saliency maps exhibit consistent overlap across utterances generated by the same model despite speaker and content variations, while salient regions differ noticeably across generators. This suggests that the attributor relies on stable generator-specific cues rather than solely on input-dependent variations.

\begin{figure}[h]
    \centering
    \includegraphics[width=0.9\linewidth]{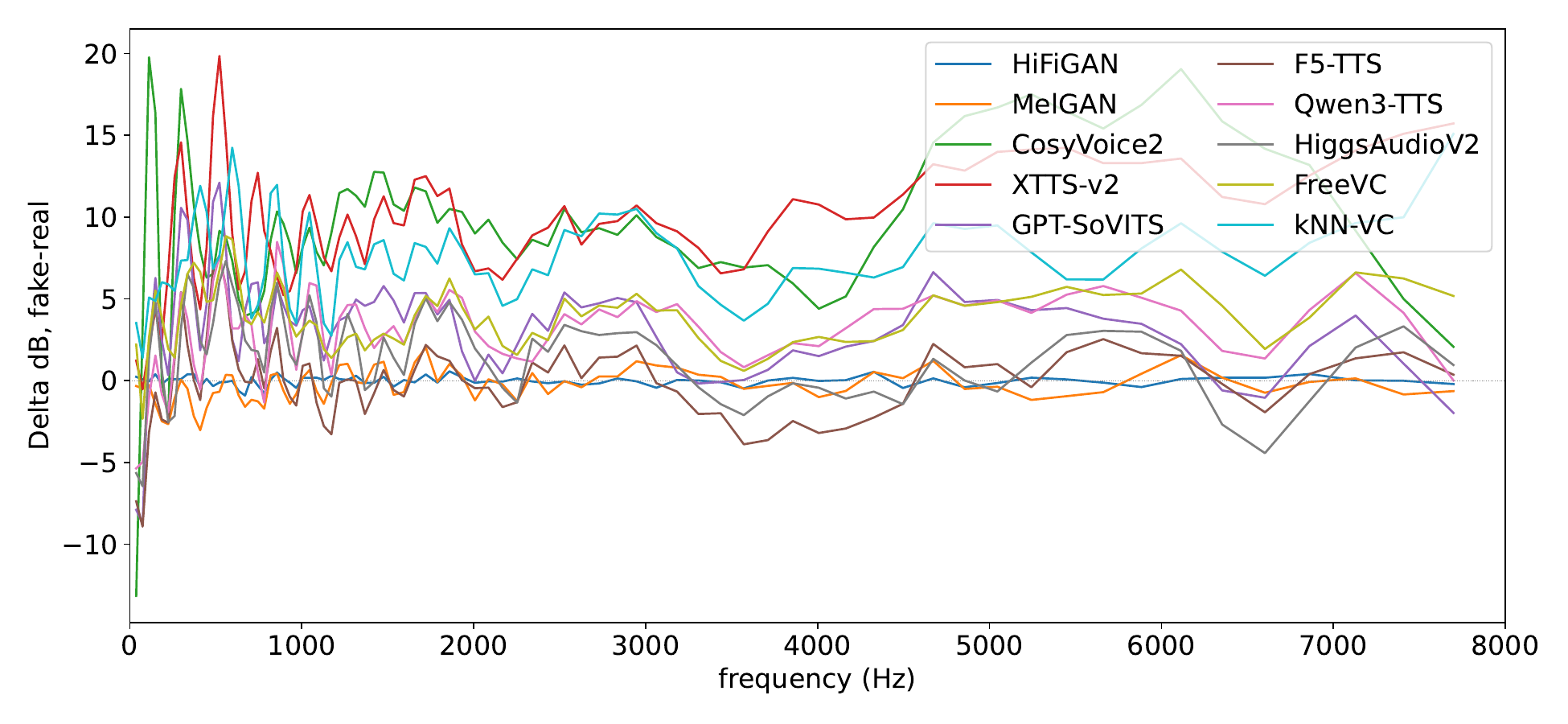}
    \caption{$\Delta E(t)$ as cross-generator distinguishability as evidence, for all 10 generators on a single held-out demo utterance (speaker p312, text T201). Positive $\Delta$ indicates the generator has more energy than the real speech in that band, negative indicates less.}
    \label{fig:cross-gen}
\end{figure}

To examine cross-generator distinguishability directly, we extend the residual-energy trajectory in Fig.~\ref{fig:cross-gen} by overlaying $\Delta E(t)$
curves for all ten generators in a single plot, each rendered in a distinct color. It shows each generator traces a visually separable spectral profile.

To quantify reproducibility, we regenerate 10 correctly attributed utterances 10 times with different random seeds under identical conditions. Deterministic vocoders show zero degradation, while stochastic generators remain highly stable with negligible accuracy and target-class probability changes. Only Qwen3-TTS and XTTS-v2 exhibit tiny measurable drops, with accuracy/probability decreases of 2.00\%/1.78\% and 6.00\%/3.65\%, respectively, indicating that attribution is largely reproducible across repeated generations.

\subsection{Audio perturbation}

We performed the perturbation experiments on the VCTK test set discussed above with results shown in Table \ref{tab:perturb_quality}. Specifically, perturbations with $\Delta$DNSMOS below 0.8 and STOI above 0.95 are regarded as perceptually transparent, where the exact threshold is adopted as an operational definition following prior work \cite{reddy2021dnsmos,taal2010short}.

\begin{table}[h]
\centering
\caption{Effect of different perturbations on attribution performance and perceptual transparency. Acc./probability drops are reported for WavLM and w2v-bert, respectively. Transparent ratio denotes the percentage of samples with $\Delta$DNSMOS $<$ 0.8 and STOI $>$ 0.95. Last column reports the standard deviation of the attribution accuracy drop across the ten generators with the w2v-bert attributor. $*$ indicate statistical significance at p$<$0.001 based on McNemar's test comparing pre- and post-perturbation attribution correctness.}
\label{tab:perturb_quality}
\resizebox{\columnwidth}{!}{
\begin{tabular}{llccccccccc}
\toprule
& & \multicolumn{2}{c}{Acc.\ Drop (\%)($\uparrow$)}
& \multicolumn{2}{c}{Prob.\ Drop (\%)($\uparrow$)}
& \multicolumn{3}{c}{Perceptual Quality}
& \multirow{2}{*}{Acc. Drop Std.} \\
\cmidrule(lr){3-4}
\cmidrule(lr){5-6}
\cmidrule(lr){7-9}
Perturbation & Parameter
& WavLM & w2v-bert
& WavLM & w2v-bert
& $\Delta$DNSMOS ($\downarrow$) & STOI ($\uparrow$) & TR (\%) ($\uparrow$)
& \\
\midrule

\multirow{2}{*}{PP}
& $\sigma=0.2$
& 11.3$^*$ & 0.8 & 10.4 & 0.8
& 0.231 & 0.9991 & 96.4
& 1.0 \\
& $\sigma=0.4$
& 19.4$^*$ & 2.4$^*$ & 18.7 & 1.9
& 0.291 & 0.9975 & 95.2
& 4.1 \\
\midrule

\multirow{2}{*}{HD}
& $r=20$ dB
& 1.7$^*$ & 0.8 & 1.1 & 0.3
& \textbf{0.040} & \textbf{0.9996} & \textbf{99.8}
& 1.3 \\
& $r=10$ dB
& 2.9$^*$ & 1.3 & 2.3 & 0.8
& 0.066 & 0.9996 & 99.4
& 2.4 \\
\midrule

\multirow{2}{*}{ES}
& $k=5$
& 41.9$^*$ & 6.8$^*$ & 40.9 & 6.3
& 0.702 & 0.9800 & 73.0
& 10.0 \\
& $k=10$
& 48.3$^*$ & 20.7$^*$ & 47.0 & 20.4
& 1.333 & 0.9283 & 2.0
& 23.2 \\
\midrule

\multirow{2}{*}{BD}
& $b=10$
& 29.9$^*$ & 48.2$^*$ & 28.9 & 47.5
& 0.205 & 0.9909 & 94.4
& 30.3 \\
& $b=8$
& \textbf{55.4$^*$} & \textbf{70.5$^*$} & \textbf{54.1} & \textbf{69.8}
& 0.681 & 0.9739 & 54.8
& 28.4 \\
\midrule

\multirow{2}{*}{BN}
& $Q=40$
& 19.7$^*$ & 5.2$^*$ & 18.9 & 4.8
& 0.081 & 0.9943 & 99.9
& 11.0 \\
& $Q=20$
& 17.5$^*$ & 4.7$^*$ & 16.4 & 4.2
& 0.094 & 0.9906 & 99.8
& 10.1 \\

\bottomrule
\end{tabular}
}
\end{table}

We validated the transparency criterion through a human listening study with 20 participants (aged 20--25 years; balanced gender). Participants judged perceptual differences between original and perturbed versions of the same 40 utterances (four per perturbation), achieving substantial agreement (Fleiss' $\kappa=0.82$). Majority voting identified only ES with $k$=10 as consistently perceptible, while other perturbations were predominantly judged imperceptible. This perceptible case also causes substantial attribution degradation and audio quality degradation (increased WER), as manual inspection confirms that the introduced noise overwhelms generator-specific cues.

More interestingly, substantial attribution degradation is observed even under largely imperceptible perturbations. For example, moderate BD ($b$=10) causes accuracy drops of 29.9\% and 48.2\% for WavLM-AASIST and w2v-bert-MLP attributors, respectively, with both reductions passing McNemar's test \cite{mcnemar1947note}. The large variance of degradation across generators provides evidence that the perturbation affects generator-specific fingerprints rather than general acoustic patterns. Consistent degradation trends across both attribution models further suggest that the effect is not detector-specific. These findings indicate that imperceptible fingerprints can be disrupted without noticeable perceptual degradation.

\subsection{Emotion ablation}

Following the previous perturbation analysis, we investigate perceptible fingerprint variations by manipulating emotion as a controllable speech attribute as a representative. We apply prompt-driven emotion control on the OOD test set and verify the generated emotions using a pre-trained HuBERT-based emotion recognition model \cite{yang2021superb}. Accuracy drop is measured relative to the neutral emotion condition. All experiments are repeated five times. Table~\ref{tab:emotion_perturb} reports the mean results with paired $t$-tests, while Table~\ref{tab:ser_emotion} reports the average SER accuracy over the five runs.
\begin{table}[h]
\centering
\caption{Attribution degradation under emotion perturbations using the w2v-bert classifier. Results are reported as mean $\pm$ standard deviation over five runs. All non-zero accuracy drops are statistically significant ($p<0.001$, paired $t$-test). Statistical significance is assessed against the neutral condition using a paired t-test.}
\label{tab:emotion_perturb}
\resizebox{0.9\columnwidth}{!}{
\begin{tabular}{llccc}
\toprule
Dataset & Generator & Emotion & Acc. Drop (\%) & Prob. Drop (\%) \\
\midrule

\multirow{6}{*}{VCTK}
& \multirow{3}{*}{HiggsAudioV3}
& Happy & 0.14 $\pm$ 0.03 & 0.13 $\pm$ 0.03 \\
& & Sad   & 0.00 $\pm$ 0.00 & 0.00 $\pm$ 0.00 \\
& & Angry & \textbf{4.00 $\pm$ 0.04 }& \textbf{4.45 $\pm$ 0.05} \\
\cmidrule{2-5}
& \multirow{3}{*}{CosyVoice2}
& Happy & 0.00$\pm$ 0.01 & 0.16$\pm$ 0.02 \\
& & Sad   & \textbf{1.00 $\pm$ 0.03} & \textbf{1.22 $\pm$ 0.03} \\
& & Angry & 0.43 $\pm$ 0.03 & 0.48 $\pm$ 0.03 \\
\midrule

\multirow{6}{*}{ESD}
& \multirow{3}{*}{HiggsAudioV3}
& Happy & 11.90 $\pm$ 0.05 & 11.65 $\pm$ 0.05 \\
& & Sad   & 1.22 $\pm$ 0.03 & 1.33 $\pm$ 0.03 \\
& & Angry & 18.45 $\pm$ 0.04 & 18.97 $\pm$ 0.05 \\
\cmidrule{2-5}
& \multirow{3}{*}{CosyVoice2}
& Happy & -0.60 $\pm$ 0.02 & -0.46 $\pm$ 0.02 \\
& & Sad   & 0.52 $\pm$ 0.03 & 0.63 $\pm$ 0.03 \\
& & Angry & -0.39 $\pm$ 0.02 & -0.30 $\pm$ 0.02 \\

\bottomrule
\end{tabular}}
\end{table}
\begin{table}[h]
\centering
\caption{Emotion recognition accuracy of generated speech evaluated by SER models.}
\label{tab:ser_emotion}
\resizebox{0.7\columnwidth}{!}{
\begin{tabular}{llccc}
\toprule
Dataset & Generator & Happy & Sad & Angry \\
\midrule
\multirow{2}{*}{VCTK}
& HiggsAudioV3 & 20.29\% & 0.71\% & 99.57\% \\
& CosyVoice2  & 32.86\% & 0.00\% & 72.71\% \\
\midrule
\multirow{3}{*}{ESD}
& Original     & 24.2\% & 0.1\% & 98.1\% \\
& HiggsAudioV3 & 13.10\% & 3.4\% & 99.7\% \\
& CosyVoice2   & 41.3\% & 0.0\% & 74.7\% \\
\bottomrule
\end{tabular}}
\end{table}

Perceptible fingerprints cause quite small attribution degradation. Emotion-based perturbations reduce attribution by less than 4.00\% for HiggsAudioV3 and around 1.00\% for CosyVoice2. This suggests that perceptible fingerprints are more readily optimized during generation, but less likely stable for attribution.

The relatively low SER accuracy does not necessarily indicate poor emotion generation, as cross-domain SER remains highly challenging, with around 40\% macro accuracy considered competitive~\cite{parry2019analysis}. To further validate emotional expressiveness, the same 20 annotators labelled 200 generated samples (50 per emotion). Majority voting achieved recognition rates of 100\% for angry, 95\% for neutral, 80\% for happy, and 78\% for sad, confirming that the intended emotions were successfully conveyed. 

\begin{figure}
    \centering
    \includegraphics[width=0.9\linewidth]{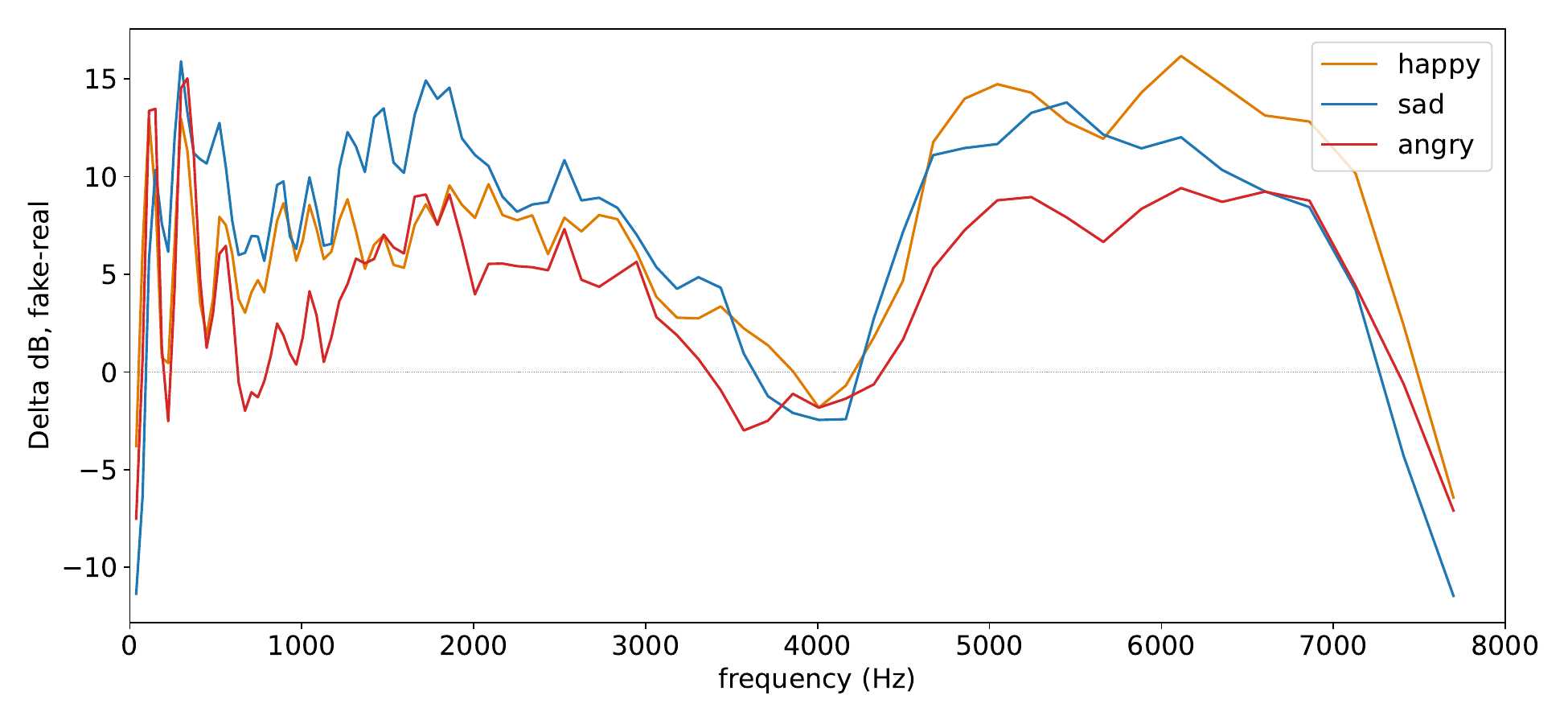}
    \caption{$\Delta E(t)$ between different emotions as evidence of imperceptible fingerprints for CosyVoice2 on a single held-out demo utterance (speaker p312, text T201).}
    \label{fig:delta_e}
    \vspace{-0.7cm}

\end{figure}

Compared with the previous Figure \ref{fig:cross-gen}, the $\Delta E(t)$ patterns under different emotion conditions remain highly consistent shown in Figure \ref{fig:delta_e}, with similar magnitudes and unchanged temporal trajectories. This indicates that the imperceptible fingerprints are preserved across emotion transformations and remain stable.

\subsection{Validity of the protocol}
Following the definition of fingerprints as generator-dependent traces left in audio, attribution changes indicate that fingerprint components are affected. If a perturbation causes generator-dependent attribution changes while remaining perceptually transparent, it reveals changes in imperceptible fingerprint components; otherwise, it reflects perceptible fingerprints. In our experiments, the large variation in attribution degradation across generators indicates that the perturbations affect generator-specific fingerprints rather than general acoustic patterns.

The protocol is considered successful as Probe 1 confirms fingerprint consistency and reproducibility, Probe 2 shows that persistent fingerprints resides in imperceptible components through substantial degradation under imperceptible perturbations, and Probe 3 shows that perceptible fingerprints are easily altered and less stable.

\section{Conclusion}

We proposed PIPDF, a diagnostic framework that separates perceptible and imperceptible fingerprints in speech deepfake detection. By combining controlled interventions with attribution analysis, PIPDF provides a systematic way to assess fingerprint reliability, revealing that imperceptible fingerprints are generally more persistent. Future work could explore dedicated datasets to assess whether imperceptible fingerprints can be mitigated through targeted optimisation.

\bibliographystyle{IEEEbib}
\bibliography{strings,refs}

\end{document}